\documentclass[aps, prl,  twocolumn,preprintnumbers,amsmath,amssymb,floatfix]{revtex4}

\usepackage{graphicx}
\usepackage{epstopdf}
\usepackage{dcolumn}
\usepackage{bm}

\begin{document}

\title{Remarks on Flory theory of a self-avoiding chain under cylindrical confinement}

\author{Suckjoon Jun}
 \email{suckjoon.jun@necker.fr}
 \affiliation{INSERM, Unit\'{e} 571, Fac Med Necker, F-75730 Paris 15, France}
 \author{Bae-Yeun Ha}
  \email{byha@uwaterloo.ca}
  \affiliation{Department of Physics and Astronomy, University of Waterloo, Waterloo, Ontario N2L 3G1, Canada}

\date{\today}

\begin{abstract}
\end{abstract}

\maketitle

Much progress in understanding chain molecules often thrives on simplification, due to their intrinsic complexity.  Nevertheless, it is hard to exaggerate the success and the impact of Flory's brilliant scheme for computing the equilibrium size of a swollen polymer chain in a good solvent~\cite{Flory}.  
Consider such a swollen polymer carrying $N$ monomers with its end-to-end distance $R$.
The {\it trial} Flory free energy in $d$ spatial dimensions is then expressed
as~\cite{endnote}  
\begin{equation}
\label{eq:Flory}
\beta \mathcal{F}_d (R) \sim \frac{R^2}{N a^2}+  \frac{a^d N^2}{R^d},
\end{equation}
where $a$ the monomer size, $\beta =1/k_BT$ with $k_B$ the Boltzmann constant and $T$ the absolute temperature.  (Here and below, we only consider an {\it athermal} case, {\i.e.}, the excluded volume of each monomer is $a^d$, independent of $T$.)   
The first term describes chain elasticity of entropic spring,
while the second term represents the meanfield energy of (two-body) interaction between monomers along the chain~\cite{Flory,deGennesBook}. 
The equilibrium chain size or the Flory radius, $R_F$, is obtained by minimizing $\mathcal{F}$ with respect to $R$:  For $d \le 4$, $R_F \sim a N^\nu$ with $\nu = 3/(2+d)$ (e.g., $\nu=3/5$ for $d=3$). This exponent $\nu$ is rightly designated as the Flory exponent.

As de Gennes correctly pointed out, however, Flory's theory benefits from the remarkable cancellation of errors (overestimates) of both terms in Eq.~\ref{eq:Flory}~\cite{deGennesBook}. Note that, at $R=R_F$, the Flory free energy scales as 
\begin{equation}
\label{eq:FreeEnergy}
\beta \mathcal{F}_d(R_F) \sim  N^{(4-d)/(2+d)}.
\end{equation}
This result is equivalent to stating that the number of monomer contacts  scales as $N^2/R_F^d \sim N^{2-\nu d}$  ($\sim N^\frac{1}{5}$ for $d=3$~\cite{FloryKrigbaum}).  For $d=3$, this is an overestimate in view of the more elaborate analysis in Ref.~\cite{Grosberg,GrosbergBook}. 
On the other hand, $\beta {\cal F}_{d=4} (R_F)\sim 1$ is expected to be asymptotically valid.

Despite the aforementioned limitations, mainly due to its simplicity, Flory-type approaches have been extended to many other important cases, e.g., linear chains with stiffness~\cite{Schaefer, Nakanishi} and polymers of various topology in a confined space (for a review, see Ref.~\cite{Vilgis00}). In particular, the widely-used free energy of a linear chain in a cylindrical pore has the following form~\cite{Brochard90, Vilgis00}
\begin{equation}
\label{eq:Vilgis}
\beta \mathcal{F}(R_\parallel, D) \sim \frac{R_\parallel^2}{N a^2}+\frac{a^3 N^2}{D^2 R_\parallel} 
,\end{equation}
where $R_\parallel$ is the trial chain size in the longitudinal direction and $D$ the width of the cylinder.  

Although Eq.~\ref{eq:Vilgis} produces the correct scaling for the equilibrium chain size $R_{\parallel 0} \sim Na (a/D)^\frac{2}{3}$~\cite{deGennesBook}, its applicability beyond the computation of $R_{\parallel 0}$ is severely limited by the following two important errors: (i) For a chain in equilibrium ($R_\parallel = R_{\parallel 0}$), Eq.~\ref{eq:Vilgis} predicts  $\beta \mathcal{F} \sim N (a/D)^{\frac{4}{3}}$, but the correct  free energy should scale as  $\beta \mathcal{F} \sim N (a/D)^\frac{5}{3}$~\cite{deGennesBook,Brochard77}.
(ii) Eq.~\ref{eq:Vilgis} results in an incorrect effective ``Hookian'' spring constant $k_\mathrm{eff}$ of the chain, $\beta k_\mathrm{eff} \sim N^{-1}a^{-2}$, but the correct scaling is $\beta k_\mathrm{eff} \sim N^{-1} D^{-\frac{1}{3}}a^{-\frac{5}{3}}$~\cite{Brochard77, Austin}.   The two errors are not unrelated, and, in fact, arise from the same source.  For instance, the entropy term of Eq.~\ref{eq:Vilgis} fails to correctly take into account the additional length scale $D$ associated with the cylinder -- it erroneously assumes that the elasticity is not affected by the presence of confinement.

Here, we propose the following ``renormalized'' free energy for a polymer under cylindrical confinement.
\begin{equation}
\label{eq:freeBY}
\beta\mathcal{F}_\mathrm{cyl}(R_\parallel,D) \sim   \frac{R_\parallel^2}{(N/g)D^2}+ \frac{D(N/g)^2}{R_\parallel} ,
\end{equation}
where $g$ is the number of monomers inside a blob of diameter $D$, {\i.e.}, $g \simeq (D/a)^\frac{5}{3}$.  Our basic idea is to consider the confined space inside a cylinder as an effective one-dimensional space ($d=1$) and, hence, to introduce the lengthscale $D$ accordingly in Eq.~\ref{eq:Flory} by rescaling $a \rightarrow a^\prime = D$ and $N \rightarrow N^\prime = N/g$.   
Then, the first term can be understood as the chain being made of $N/g$ subunits (``blobs'') of size $D$, while the second term describes the mutual exclusion between neighboring blobs. 
In other words, Eq.~\ref{eq:freeBY} represents a self-avoiding chain confined in a cylinder of diameter $D$ as a one-dimensional Rouse (or Gaussian) chain with an effective step length $D$, since long-distance interactions beyond $D$ (along the contour) are suppressed; thus, it is also expected to 
exhibit the Rouse dynamics
~\cite{Brochard77}.

Indeed, the above free energy produces not only the expected equilibrium chain size $R_{\parallel 0} \sim Na (a/D)^\frac{2}{3}$, but also the correct blob-overlapping free energy of  
$\beta \mathcal{F}_\mathrm{conf} \sim N/g \sim N (a/D)^\frac{5}{3}$ (namely, the total number of blobs)~\cite{deGennesBook,Brochard77}.  Note that this is identical to confinement free energy.  The way neighboring blobs interact each other is not different from the way they are constrained by the confining wall.  Importantly, we obtain the correct effective Hookian spring constant of the chain: $\beta k_\mathrm{eff} \sim \partial^2 \mathcal{F}_\mathrm{cyl}/\partial R_\parallel^2 |_{R_\parallel = R_{\parallel 0}}  \sim N^{-1} a^{-\frac{1}{\nu}}D^{\frac{1}{\nu}-2}\sim N^{-1}a^{-2}(a/D)^\frac{1}{3}$~\cite{Brochard77,Austin}. Moreover, using the stretch-release argument,  the global (slowest) relaxation time of the confined chain (in the absence of hydrodynamic effects) is then reciprocally obtained as $\tau_\mathrm{R} \sim N/k_\mathrm{eff} \sim N^2 a^\frac{1}{\nu}D^{2-\frac{1}{\nu}} \sim N^2 a^2 (D/a)^\frac{1}{3} $, consistent with the earlier scaling result~\cite{Brochard77, Austin}.

A natural extension of Eq.~\ref{eq:freeBY} to the slit case~\cite{Rubinstein} would be obtained from Eq.~\ref{eq:Flory} with the same rescaling of  $a \rightarrow a^\prime = D$ and $N \rightarrow N^\prime = N/g$, assuming $d=2$, as follows
\begin{equation}
\label{eq:freeslit}
\beta\mathcal{F}_\mathrm{slit}(R_\parallel,D) \sim   \frac{R_\parallel^2}{(N/g)D^2}+ \frac{D^{2}(N/g)^2}{R_\parallel^{2}},
\end{equation}
where $D$ is the distance between the two parallel slits and $g \sim (D/a)^{5/3}$.  Indeed, this free energy predicts the correct equilibrium size of the chain in a slit, $R_{\parallel 0}  \sim a N^{3/4} (a/D)^{1/4}$, in agreement with the results of other approaches~\cite{deGennesBook, Vilgis00}. However, its equilibrium free energy, $\beta \mathcal{F}_\mathrm{slit}(R_\parallel = R_{\parallel 0}, D)$, scales as  $ N^{1/2} (a/D)^\frac{5}{3}$. Note that this is different from the free energy of slit confinement $\beta \mathcal{F}_\mathrm{conf} \sim N/g \sim N (a/D)^\frac{5}{3}$, which is identical to that of a cylinder
(see Ref.~\cite{deGennesBook, Rubinstein} and references therein). 
The slit free energy in Eq.~\ref{eq:freeslit} describes the arrangement of blobs in a slit, not confinement.

Why does the Flory approach work better for the cylinder case than for higher dimensions? 
To see the difference, consider the volume fraction ($\alpha$) of monomers inside a space explored by a self-avoiding chain in equilibrium. For cylindrical confinement, this is given by $\alpha = Na^3 /D^2 R_{\parallel 0} \sim (D/a)^{-4/3}$, \emph{independent} of $N$, whereas, both for a slit and for a dilute bulk solution, $\alpha \rightarrow 0$ as $N \rightarrow \infty$.  It is this special nature of compactness of one-dimensional space that explains the extensiveness and thus the success
of the rescaled Flory approach in Eq.~\ref{eq:freeBY}~\cite{endnote2}.

Extension of our analysis to semi-flexible polymers and chains with non-trivial topology (e.g., branched) is important for understanding many recent experiments on DNA in nano- and micro-channels~\cite{Craighead} and have non-trivial implications for practical applications such as filtration~\cite{Sakaue}.
          
We thank Daan Frenkel for helpful discussion.  We acknowledge financial supports from NSERC (Canada, BYH) and  NSERC post-doctoral fellowship (Canada, SJ).


\begin{thebibliography}{19}
\bibitem{endnote2}
We also note that this compactness of the 1-$d$ case agrees with our meanfield picture (Cf. Eq.~\ref{eq:freeBY}).  To see this, we calculate the relative end-to-end distance fluctuation of the chain using Eq.~\ref{eq:freeBY}, $\delta R_\parallel/R_{\parallel 0}\sim {k_\mathrm{eff}}^{-\frac{1}{2}}R_{\parallel 0}^{-1} \sim (D/a)^\frac{11}{6}N^{-\frac{1}{2}}$.  (This is consistent with the scaling analysis~\cite{Brochard77}).  Thus, for $(D/a)^\frac{11}{3} \ll N$, the overall chain fluctuations become negligible.  On the other hand, the non-compactness of higher spatial dimensions makes the effect of fluctuations more serious in that any meanfield approach is less reliable. For instance, Eq.~\ref{eq:Flory} for $d=3$ erroneously implies that $\delta R/R_F \sim N^{-\frac{1}{10}} \rightarrow 0$ for large $N$.  The correct relation is $\delta R  \sim R_F$ -- the chain size in this compact non-compact  case is set by the end fluctuation.  We believe Eq.~\ref{eq:freeslit} suffers from a similar inconsistency.
\end{thebibliography}
\end{document}